# Strong and Tunable Electrical-Anisotropy in Type-II Weyl Semimetal Candidate $WP_2$ with Broken Inversion Symmetry

*Bo Su[†], Yanpeng Song[†], Yanhui Hou[†], Xu Chen, Jianzhou Zhao, Yongchang Ma, Yang Yang, Jiangang Guo, Jianlin Luo, and Zhi-Guo Chen**

B. Su, Y. Song, Y. Hou, X. Chen, Prof. Y. Yang, Prof. J. Guo, Prof. J. Luo, Prof. Z.-G. Chen
Beijing National Laboratory for Condensed Matter Physics
Institute of Physics, Chinese Academy of Sciences
Beijing 100190, China
Email: zgchen@iphy.ac.cn

B. Su, Y. Song, X. Chen, Prof. J. Guo, Prof. Z.-G. Chen
School of Physical Sciences
University of Chinese Academy of Sciences
Beijing 100190, China

Y. Hou, Prof. Y. Ma
School of Materials Science and Engineering
Tianjin University of Technology
Tianjin 300384, China

Prof. J. Zhao
Co-Innovation Center for New Energetic Materials
Southwest University of Science and Technology
Mianyang, Sichuan 621010, China

Prof. Y. Yang, Prof. J. Guo, Prof. J. Luo, Prof. Z.-G. Chen
Songshan Lake Materials Laboratory
Dongguan, Guangdong, 523808, China

Prof. J. Luo
Collaborative Innovation Center of Quantum Matter
Beijing, China

Keywords: type-II Weyl semimetal, in-plane electrical anisotropy, optical spectroscopy, Raman spectroscopy, first-principle calculations

**A transition metal diphosphide, $WP_2$, is a candidate for type-II Weyl semimetals (WSMs) in which spatial inversion symmetry is broken and Lorentz invariance is violated. As one of the key prerequisites for the presence of the WSM state in $WP_2$, spatial inversion symmetry breaking in this compound has rarely been investigated by experiments. Furthermore, how much anisotropy the electrical properties of $WP_2$ have and whether its electrical anisotropy can be tuned remain elusive. Here, we report angle-resolved polarized Raman spectroscopy, electrical transport, optical spectroscopy and first-principle studies of $WP_2$. The energies of the observed Raman-active phonons and the angle dependences of the detected phonon intensities are well consistent with the results obtained by first-principle calculations and the analysis of the proposed crystal symmetry without spatial inversion, providing evidence that spatial inversion symmetry is *broken* in $WP_2$. Moreover, the measured ratio ($R_c/R_a$) between the crystalline *c*-axis and *a*-axis electrical resistivities exhibits a weak dependence on temperature ($T$) in the temperature range from 100 to 250 K, *but increases abruptly* at $T \leq 100$ K, and then reaches the value of ~ 8.0 at $T = 10$ K, which is by far the *strongest* in-plane electrical resistivity anisotropy among the reported type-II WSM candidates with comparable carrier concentrations. Our optical spectroscopy study, together with the first-principle calculations on the electronic band structure, reveals that the abrupt enhancement of the electrical resistivity anisotropy at $T \leq 100$ K mainly arises from a sharp increase in the scattering rate anisotropy at low temperatures. More interestingly, the $R_c/R_a$ of $WP_2$ at $T = 10$ K can be *tuned* from 8.0 to 10.6 as the magnetic field increases from 0 to 9 T. The so-far-strongest and magnetic-field-tunable electrical resistivity anisotropy found in $WP_2$ can serve as a degree of freedom for tuning the electrical properties of type-II WSMs, which paves the way for the development of novel electronic applications based on type-II WSMs.**

A Weyl semimetal (WSM), which is a type of quantum matters with pairs of Weyl cones in the bulk and topologically protected Fermi arcs on the surface,[1-10] offers an excellent platform for the realization of exotic phenomena, such as the chiral magnetic effect,[11-17] high efficient catalysis [18,19] and quantum anomalous Hall effect.[20] Two types of WSMs can be achieved by breaking spatial inversion symmetry or time reversal symmetry.[1-10,20-27] Type-I WSMs can have closed point-like Fermi surfaces in the bulk, while type-II WSMs host open bulk Fermi surfaces and Weyl points at the boundary between electron and hole pockets.[28-33] Furthermore, type-II WSMs are expected to possess unusual transport and optical properties which are distinct from those of type-I WSMs.[34-37] However, to date, only a few type-II WSMs have been identified experimentally. A natural question to ask is whether a broader class of type-II WSMs with novel physical properties, such as strong and tunable electrical anisotropy, can be found experimentally.

Recently, a transition metal diphosphide, $WP_2$, was theoretically predicted to be a candidate for type-II WSMs.[32] Four pairs of Weyl points with opposite chirality are anticipated to be located below the Fermi level of $WP_2$. Therein, the distance between the two Weyl points of each pair in momentum space is so long that the nearest Weyl points have the same chirality. Since (i) the positions of the Weyl points are sensitive to the small perturbations of the crystal structure, such as structural distortions and defects, and (ii) the two Weyl points with opposite chirality would annihilate if they meet in momentum space, the same chirality of the neighboring Weyl points makes the predicted type-II WSM state of $WP_2$ particularly robust against structural distortions or defects. This distinguished character—the protection of the neighboring Weyl points with the same chirality against annihilation upon structural distortions or defects significantly reduces the probability of the scattering between the neighboring Weyl points and thereby plays an important role in the extremely low resistivity observed in $WP_2$.[38-40] In addition, the thermal and electrical transport measurements revealed that the $WP_2$ single crystals exhibit the transport behaviors of a hydrodynamic electron fluid: a viscosity-induced dependence of the electrical resistivity on the sample width and a strong violation of the Wiedemann–Franz law (i.e., the product of the thermal conductivity and the electrical resistivity, divided by the temperature is the Sommerfeld value which depends only on fundamental constants).[41-43] However, to date, the anisotropy degree of the electrical resistivity within the crystalline *ac*-plane of $WP_2$ has not been explored by experiments.

Furthermore, whether the anisotropy of the *ac*-plane resistivity of $WP_2$ can be tuned is still unclear. Generally, the electrical properties of solids are intimately associated with the crystal structures. Thus, investigating the crystal symmetry of $WP_2$ would be helpful in exploring and understanding the anisotropy degree of its electrical resistivity. Nevertheless, compared with the phosphorus positions of $MoP_2$ with non-centrosymmetric $Cmc2_1$ space group, the phosphorus positions of $WP_2$, which are crucial for deriving its crystal symmetry, are less accurately determined by X-ray diffraction due to the larger discrepancy between the electron-density distributions of the transition-metal and phosphorus atoms. Moreover, for nonmagnetic $WP_2$, the predicted Fermi arcs on the surface, which is a critical feature of a nonmagnetic Weyl semimetal with broken spatial inversion symmetry, has so far not been observed clearly by angle-resolved photoemission spectroscopy.[38,44,45] Therefore, although $WP_2$ was proposed to be a promising type-II WSM candidate with remarkable properties, such as extremely low resistivity and hydrodynamic transport behaviors, one of the key prerequisites for the presence of the intriguing WSM state in $WP_2$—spatial inversion symmetry breaking in this compound needs to be further confirmed.

Angle-resolved polarized Raman spectroscopy (ARPRS) is an experimental technique which is directly sensitive to the crystal symmetry. However, there are few ARPRS studies of $WP_2$. Thus, a systematic investigation on the polarized Raman spectra of $WP_2$ is essential to fill the gap between the theoretical expectation of the spatial inversion symmetry breaking and the experimental confirmations. Here, we use ARPRS to check whether spatial inversion symmetry is broken in $WP_2$. The measured Raman-active phonon energies and the angle dependences of the detected phonon intensities agree well with the results obtained by first-principle calculations and the analysis of the proposed $Cmc2_1$ space group without spatial inversion. The good consistence between the experimental Raman data and theoretical results indicates spatial inversion symmetry breaking in $WP_2$. The angle-dependent Raman-active phonon peak intensities not only indicate that the Raman-active phonon (or lattice) vibrations are anisotropic in $WP_2$, but also manifest that ARPRS is a rapid and nondestructive tool to determine the crystallographic orientation of anisotropic type-II WSM candidates including $WP_2$. Since the lattice structure of a material has a close relationship with its electrical properties, to investigate the degree of the electrical anisotropy, we carried out electrical resistivity measurements of the $WP_2$ single crystals at different temperatures and at different magnetic fields. Here, the ratio (i.e., $R_c/R_a$) between the electrical resistivity measured with the electric current (i.e., $\boldsymbol{I}$) along the crystalline *c*-axis and the electrical resistivity measured with $\boldsymbol{I}$ // a-axis exhibits a weak dependence on temperature ($T$) (i.e., $R_c/R_a \sim 1.6$) in the temperature range from 100 to 250 K, *but increases abruptly* at $T \leq 100$ K, and then reaches the value of ~ 8.0 at $T = 10$ K, which is by far the *strongest* electrical resistivity anisotropy among the reported type-II WSM candidates with their carrier concentrations comparable to ~ $10^{21}$ $cm^{-3}$. Because optical spectroscopy is an efficient experimental tool for studying charge dynamics in materials, to study the origin of the abrupt increase in the resistivity anisotropy at low temperatures, we performed optical reflectance measurements on the $WP_2$ single crystals at different temperatures with the electrical field of the incident light applied along the crystalline *c*-axis and *a*-axis, respectively. Our optical spectroscopy study, together with the first-principle calculations on the electronic band structure, indicates that the steep enhancement of the electrical resistivity ratio $R_c/R_a$ largely comes from a sharp increase in the scattering rate anisotropy at low temperatures. More interestingly, the $R_c/R_a$ of $WP_2$ at $T = 10$ K can be tuned from 8.0 to 10.6 as the magnetic field ($B$) applied perpendicular to the crystalline *ac*-plane increases from 0 to 9 T. The strong and tunable anisotropy (1060% at $B$ = 9 T and 800% at $B$ = 0 T) of the in-plane (i.e., *ac*-plane) electrical resistivity can be considered as a degree of freedom for tuning the electrical properties of type-II WSMs.

$WP_2$ was proposed to crystallize in the space group of $Cmc2_1$ and has a non-centrosymmetric structure with a two-fold screw axis along the crystalline *c*-axis, a glide plane perpendicular to the *b*-axis and a mirror plane perpendicular to the *a*-axis,[38] shown in **Figure** 1a. To study whether spatial inversion symmetry is broken in the crystal structure of $WP_2$, we firstly used a 532 nm wavelength laser to measure the Raman spectra of the $WP_2$ single crystals at room temperature in the parallel-polarized configuration (i.e., the electrical field of the linearly polarized incident light $\boldsymbol{E_i}$ is parallel to the electrical field of the linearly polarized scattered light $\boldsymbol{E_s}$, here $\boldsymbol{E_i}$ // $\boldsymbol{E_s}$ // *a*-axis) and in the perpendicular-polarized configuration (i.e., $\boldsymbol{E_i} \perp \boldsymbol{E_s}$, here $\boldsymbol{E_i}$ // *a*-axis and $\boldsymbol{E_s}$// *c*-axis), respectively (see the schematic of the configurations of the ARPRS in **Figure** 1b, the crystal orientation of the $WP_2$ single crystal characterized by single-crystal X-ray diffraction in the bottom-left inset of **Figure** 1b and the details about growing the single crystals, measuring the Raman spectra and determining the crystal orientation in Experimental Section). **Figure** 1c displays the polarized Raman spectra of the $WP_2$ single crystals in the perpendicular- and parallel-polarized configurations. Three peak-like features can be observed in the perpendicular-polarized configuration, while eight peak-like features are present in the parallel-polarized configuration. To study the nature of the probed peak-like features, we performed the analysis of the proposed $Cmc2_1$ space group. Since (i) the proposed $Cmc2_1$ space group belongs to the $C_{2V}$ point group, and (ii) the 6 atoms of its unit cell correspond to 18 phonon modes, the phonon modes of $WP_2$ should be decomposed into 18 irreducible representations: $[5A_1 + 3A_2 + 2B_1 + 5B_2] + [5A_1 + 2B_1 + 5B_2] + [A_1 + B_1 + B_2]$, where the first, second and third terms represent the Raman-active, infrared-active and the acoustic phonon modes, respectively (see the elaboration on the derivation of the irreducible representations in Section 1 of Supplemental Information). Therein, the symmetry of the Raman-active phonons is determined by the corresponding Raman tensors ($\boldsymbol{R}$) of the $C_{2V}$ point group:

$$\boldsymbol{R}(A_1) = \begin{pmatrix} a & 0 & 0 \\ 0 & b & 0 \\ 0 & 0 & c \end{pmatrix}, \qquad \boldsymbol{R}(A_2) = \begin{pmatrix} 0 & d & 0 \\ d & 0 & 0 \\ 0 & 0 & 0 \end{pmatrix},$$

$$\boldsymbol{R}(B_1) = \begin{pmatrix} 0 & 0 & e \\ 0 & 0 & 0 \\ e & 0 & 0 \end{pmatrix}, \qquad \boldsymbol{R}(B_2) = \begin{pmatrix} 0 & 0 & 0 \\ 0 & 0 & f \\ 0 & f & 0 \end{pmatrix},$$

where the *a*, *b*, *c*, *d*, *e* and *f* are the tensor elements and their values depend on the cross section of Raman scattering. Moreover, the intensity ($I_{Ph}$) of the Raman-active phonon has the following relationship with the polarization states of the incident light $\boldsymbol{e_i}$ and the scattered light $\boldsymbol{e_s}$ and $\boldsymbol{R}$ [46]

$$I_{Ph} \propto |\boldsymbol{e_i} \cdot \boldsymbol{R} \cdot \boldsymbol{e_s}|^2. \tag{1}$$

On the basis of the irreducible representations of the $C_{2V}$ point group, the above Raman tensors and the above relationship among the phonon intensity, the Raman tensors and the polarization states of the incident and scattered lights, we can obtain that (i) five $A_1$ Raman-active phonon modes and three $A_2$ Raman-active phonon modes can be observed under the parallel-polarized (i.e., $\boldsymbol{E_i}//\ \boldsymbol{E_s}//\ a$-axis) and perpendicular-polarized (i.e., $\boldsymbol{E_i} \perp \boldsymbol{E_s}$ and $\boldsymbol{E_s}//\ c$-axis) configurations, respectively, and that (ii) two $B_1$ and five $B_2$ Raman-active phonon modes can be detected only when the polarized electrical field $\boldsymbol{E_i}$ or $\boldsymbol{E_s}$ is aligned along the *b*-axis which is perpendicular to the *ac*-plane. Using first-principles calculations, we got the theoretical energies of the Raman-active phonon modes in $WP_2$ (see the details about our first-principle calculations in Experimental Section). The energies of the calculated phonon modes and the measured peak-like features are listed in Table 1 (see the calculated phonon band dispersions of $WP_2$ along the high-symmetry lines in **Figure** S2b). According to the above group analysis and the consistence between the theoretical and experimental Raman-active phonon energies, (i) the three measured peak-like features, $P_1$ at 159.2 $cm^{-1}$, $P_4$ at 256.3 $cm^{-1}$, and $P_5$ at 282.2 $cm^{-1}$, which are measured in the perpendicular-polarized configuration, can be assigned to the three Raman-active phonon modes $3A_2$, and (ii) the five measured peak-like features, $P_2$ at 168.4 $cm^{-1}$, $P_6$ at 284.1 $cm^{-1}$, $P_8$ at 359.9 $cm^{-1}$, $P_9$ at 392.4 $cm^{-1}$, and $P_{10}$ at 511.7 $cm^{-1}$, which are measured in the parallel-polarized configuration, can be ascribed to the five Raman-active modes $5A_1$ (see the discussion about the three peak-like features, $P_3$, $P_7$ and $P_{11}$ in Section 2 of Supplemental Information). **Figure** 2 shows the atomic schematics of the calculated phonon vibrational patterns for the five Raman-active phonon modes $5A_1$ in the parallel-polarized configuration and the three Raman-active phonon modes $3A_2$ in the perpendicular-polarized configurations (see the vibrational patterns for the seven Raman modes, $2B_1 + 5B_2$, in **Figure** S3). Besides the non-centrosymmetric structure with the space group $Cmc2_1$ discussed above, $WP_2$ can crystallize in a centrosymmetric structure with the space group *C12/m1*, which has spatial inversion symmetry and belongs to the $C_{2h}$ point group.[47,48] After performing the analysis of the possible centrosymmetric space group *C12/m1*, we found that the phonon modes of $WP_2$ corresponding to this space group should be decomposed into 18 irreducible representations: $[6A_g + 3B_g] + [2A_u + 4B_u] + [A_u + 2B_u]$, where the first, second and third terms represent the Raman-active, infrared-active and the acoustic phonon modes, respectively. In the centrosymmetric space group of $WP_2$, the parity-even $A_g$ (or $B_g$) and the parity-odd $A_u$ (or $B_u$) representations are exclusive with each other. The Raman tensors ($\boldsymbol{R}$) of the $C_{2h}$ point group, which correspond to the $A_g$ and $B_g$ Raman-active phonon modes, can be written as:

$$\boldsymbol{R}(A_g) = \begin{pmatrix} a & d & 0 \\ d & b & 0 \\ 0 & 0 & c \end{pmatrix}, \qquad \boldsymbol{R}(B_g) = \begin{pmatrix} 0 & 0 & e \\ 0 & 0 & f \\ e & f & 0 \end{pmatrix},$$

According to the irreducible representations of the $C_{2h}$ point group, the corresponding Raman tensors and the relationship among the phonon intensity, the Raman tensors and the polarization states of the incident and scattered lights (i.e., Equation (1)), only six $A_g$ Raman-active phonon modes can be observed in the perpendicular-polarized or parallel-polarized configuration. Thus, the only two space groups (i.e., *Cmc2*$_1$ with broken inversion symmetry and *C12/m1* with inversion symmetry) of $WP_2$ have different numbers of Raman-active phonon modes in the perpendicular-polarized and parallel-polarized configuration (the consistency between the numbers of the measured Raman-active phonon modes in **Figure** S2a measured by the 532 nm wavelength laser and the 633 nm wavelength laser in the parallel-polarized configuration suggests that the assignment of the group symmetry of $WP_2$ is unlikely to be influenced by the wavelength of the laser used for the Raman measurements[49]). Moreover, our polarized Raman measurements of $WP_2$ show three peak-like features in the perpendicular-polarized configuration and eight peak-like features in the parallel-polarized configuration. Therefore, the inconsistency between the numbers of the measured peak-like features in **Figure** 1c and the theoretical Raman-active phonon modes obtained based on the centrosymmetric space group *C12/m1* with inversion symmetry indicates that the crystal structure of the $WP_2$ single crystals here is unlikely to belong to the centrosymmetric space group *C12/m1* with inversion symmetry.

To further investigate whether spatial inversion symmetry is broken in $WP_2$, we performed the crystal-angle-resolved Raman measurements at room temperature under the parallel-polarized (i.e., $\boldsymbol{E_i}$ // $\boldsymbol{E_s}$) and perpendicular-polarized (i.e., $\boldsymbol{E_i} \perp \boldsymbol{E_s}$) configurations with $\boldsymbol{E_i}$ and $\boldsymbol{E_s}$ parallel to the crystalline *ac*-plane using a 532 nm wavelength laser. The polarized Raman spectra were measured by rotating the $WP_2$ single crystal within its *ac*-plane and altering the angle $\theta$ between the rotated crystalline $a$-axis and the fixed electrical field of the linearly polarized incident light $\boldsymbol{E_i}$ (see the schematic of the experimental setup for the crystal-angle-resolved polarized Raman measurements in **Figure** 1b). **Figure** 3a and 3b depict the crystal-angle-dependent Raman spectra of the $WP_2$ single crystals measured in the parallel-polarized configuration and the corresponding contour map of the angle-dependent Raman intensity, respectively (see the angle-dependent Raman spectra in the perpendicular configuration in **Figure** S4 and the related discussion in Section 3 of Supplemental Information). The peak-like features, $P_1$, $P_2$, $P_4$, $P_6$, $P_8$, $P_9$, and $P_{10}$, which correspond to the Raman-active phonon modes $A_1$ and $A_2$, respectively, can be observed in the angle-dependent Raman spectra under both the parallel-polarized and perpendicular-polarized configurations because when the $WP_2$ single crystal is rotated within its $ac$-plane, the electrical fields of the linearly polarized incident and scattered lights, $\boldsymbol{E_i}$ and $\boldsymbol{E_s}$ can have the two components which are along the $a$-axis and $c$-axis, respectively. As shown in **Figure** 3c-i, under the parallel-polarized configuration, the angle dependences of the intensities of the $A_1$ phonon peaks, $P_2$, $P_6$, $P_8$, $P_9$, and $P_{10}$, exhibit a two-lobed shape with two maximum intensities. For the $A_1$ phonon peaks, $P_6$ and $P_{10}$ have two maximum intensities at angle $\theta$ = 90° and 270°, while $P_2$, $P_8$ and $P_9$ show two maximum intensities at angle $\theta$ = 0° and 180°. By contrast, under the parallel-polarized configuration, the angle dependences of the intensities of the $A_2$ phonon peaks, $P_1$ and $P_4$, display a four-lobed shape with four maximum intensities at angle $\theta$ = 45°, 135°, 225°, and 315°. To quantitatively check whether the variation of the $A_1$ and $A_2$ phonon peak intensities are consistent with the spatial inversion symmetry breaking in the $WP_2$ crystal structure, we need to calculate the angle dependences of the $A_1$ and $A_2$ phonon peak intensities based on Equation (1). Under the parallel-polarized configuration, the polarization states of the incident $\boldsymbol{e_i}$ and the scattered light $\boldsymbol{e_s}$ can be expressed as $\boldsymbol{e_i}$ = $\boldsymbol{e_s}$ = (cos $\theta$, sin $\theta$, 0). According to Equation (1), the intensities of the $A_1$ and $A_2$ phonon peaks are given by [50-53]

$$I_{A_1}^{//} \propto a^2\, cos^4\,\theta + b^2\, sin^4\,\theta + 2ab\, cos^2\,\theta\, sin^2\,\theta \cos \emptyset_{ab,} \tag{2}$$

$$I_{A_2}^{//} \propto d^2\, sin^2\, 2\theta, \tag{3}$$

where $\emptyset_{ab}$ is the phase difference between the Raman tensor elements $a$ and $b$. Equation (2) and (3) indicate that in the parallel polarized configurations, the $A_1$ phonon peak intensities vary with a period of 180°, while the variation of the $A_2$ phonon peak intensities has a period of 90° with the maximum intensities at angle $\theta$ = 45°, 135°, 225°, and 315°, and the minimum intensities at angle $\theta$ = 0°, 90°, 180°, and 270°. **Figure** 3c-i show that the angle dependences of the $A_1$ and $A_2$ phonon peak intensities measured in the parallel-polarized configuration can be consistently fitted by Equations (2) and (3). Moreover, the angles of the maximal intensities of these Raman-active phonon peaks measured using ARPRS are not consistent with those corresponding to the centrosymmetric space group *C12/m1*, which further indicates that the crystal structure of the $WP_2$ single crystals here should not belong to the centrosymmetric space group *C12/m1* with inversion symmetry (see the detailed discussion in Section 4 of Supplemental Information). The good agreement between the angle dependences of the Raman-active phonon peak intensities measured using ARPRS (see the red dots in **Figure** 3c-i) and the fitting results based on the broken spatial inversion symmetry of the $WP_2$ crystal structure (see the red curves in **Figure** 3c-i), combined with the consistence between the energies of the peaks observed in the polarized Raman spectra and the calculated Raman-active phonon energies, provides convincing evidence for the spatial inversion symmetry breaking in the type-II Weyl semimetal candidate $WP_2$ by a combination of linearly polarized Raman measurements and first-principle calculations.

It is worth noticing that since the naturally grown $WP_2$ single crystals sometimes have arbitrary shapes (see **Figure** S6), the crystallographic orientation of the naturally grown crystals cannot be always determined. In the parallel-polarized configuration, the variation of the $A_1$ phonon peak intensities has a variation period of 180° and the maximum intensities appear when the electrical field of the linearly polarized incident light $\boldsymbol{E_i}$ is applied along the crystalline $a$-axis or $c$-axis, while the $A_2$ phonon peak intensities vary with a period of 90°. Therefore, the $A_1$ phonon peaks can be used to determine the crystallographic orientation of the $WP_2$ single crystal. Whether the maxima of the $A_1$ phonon peak intensity correspond to the crystalline $a$-axis or $c$-axis lies with the relative magnitude of the Raman tensor elements *a* and *b* for the $A_1$ phonon peak. According to Equation (2), when the Raman tensor elements $a > b$, the maximum intensities of the $A_1$ phonon peaks point to the *a*-axis; when the Raman tensor elements $a < b$, the maximum intensities of the $A_1$ phonon peaks are along the *c*-axis. Therefore, the crystallographic orientation of the $WP_2$ single crystals can be identified by the maximum intensities of the $A_1$ phonon peaks as long as the relative magnitude of Raman tensor elements *a* and *b* has been confirmed. We experimentally obtained that in $WP_2$, the Raman tensor elements $a > b$ for the $P_2$, $P_8$ and $P_9$ peak-like features and that the Raman tensor elements $a < b$ for the $P_6$ and $P_{10}$ peak-like features using polarized Raman spectroscopy and single-crystal X-ray diffraction on the same single crystal. Thus, the crystal-angle-resolved polarized Raman spectroscopy provides a rapid and nondestructive method for determining the crystal orientation of the $WP_2$ single crystal and other type-II WSMs.[50,51,54-58]

As revealed by ARPRS, the angle-dependent Raman-active phonon peak intensities further indicate the anisotropy of the Raman-active phonon (or lattice) vibrations in $WP_2$. Usually, the lattice structures of solids are closely related to the electrical properties. Thus, the electrical properties of $WP_2$ are expected to be anisotropic. Nevertheless, the anisotropy degree of the electrical properties of $WP_2$ remains unclear. To study the anisotropy degree of the electrical properties of $WP_2$, we measured the electrical resistivities of its single crystals at different temperatures with the electric currents (i.e., $\boldsymbol{I}$) applied along the two crystalline directions—*c*-axis and *a*-axis, respectively (see the details about the electrical resistivity measurements in Experimental Section). **Figure** 4a shows the temperature (i.e., $T$) dependences of the *c*-axis and *a*-axis resistivities (i.e., $R_c$ and $R_a$) on a logarithmic scale (see the $R_c$ and $R_a$ on the original scale in the inset of **Figure** 4a). The discrepancy between the $R_c$ and the $R_a$ can be observed clearly in **Figure** 4a and its inset. To investigate the anisotropy of the in-plane (i.e., *ac*-plane) electrical resistivity of the $WP_2$ single crystals, we plotted the ratio between the $R_c$ and the $R_a$ (i.e., $R_c/R_a$) as a function of temperature in **Figure** 4b. In the temperature range from 100 to 250 K, the $R_c/R_a$ shows a weak dependence on temperature and have the value of ~ 1.6. Surprisingly, at $T \leq 100$ K, the resistivity ratio $R_c/R_a$ of $WP_2$ increases sharply and then reaches the value of ~ 8.0 at 10 K (The relative resistivity ratio $\Delta R_c/R_a$ in the inset of **Figure** 4b, which was obtained by subtracting a linear $T$ dependence of the $R_c/R_a$ shown in the temperature range from 150 to 200 K from the $R_c/R_a$, clearly shows a sharp increase at $T \leq 100$ K), which is by far the strongest in-plane electrical resistivity anisotropy among the reported type-II WSM candidates (such as $MoTe_2$ and $TaIrTe_4$) with their carrier densities comparable to ~ $10^{21}$ $cm^{-3}$.[38,50,59,60]

In order to investigate the origin of the abrupt increase in the electrical resistivity anisotropy of $WP_2$, we calculated its electronic band structure and performed optical reflectance measurements on its single crystals with the electric field (***E***) of the incident light applied along its crystalline $c$-axis and $a$-axis, respectively (see the details about the theoretical calculations and optical reflectance measurements in Experimental Section). **Figure** 4c shows the electronic band structure of $WP_2$ obtained by first-principle calculations. The upper left and right insets of **Figure** 4c display the positions of the four pairs of Weyl points in the Brillouin zone and the Fermi surfaces, respectively. According to the positions of the Weyl points (e.g., $W_1$) in the Brillouin zone, the band dispersions near the Weyl points (e.g., the band dispersion near $W_1$ in **Figure** 4c) contribute to the butterfly-like Fermi surfaces and the cylinder-like Fermi surfaces in the upper right inset of **Figure** 4c.[32,38] Since the electrical transport properties of a material are intimately associated with the electronic states near the Fermi energy ($E_F$), to study the in-plane electrical resistivity anisotropy in $WP_2$, we focus on the electronic bands which disperse parallel to the $k_x$ and $k_z$ axis, respectively, and cross the Fermi level. The calculated band dispersion along the $k_x$ axis (i.e., the high-symmetry line Γ-X of the Brillouin zone) have an effective mass $m_x$* of ~ 1.8 $m_0$ at $E_F$. Moreover, at $E_F$, the effective masses $m_z$* of the electronic bands dispersing parallel to the $k_z$ axis (i.e., the high-symmetry lines S-R and $X_1$-$A_1$) are about 1.4 $m_0$ and 1.1 $m_0$, respectively. Therefore, the ratio (i.e., $m_z^*/m_x^*$) between the effective masses of the electronic bands dispersing parallel to the $k_z$ and $k_x$ axis is smaller than unity. Considering that (i) the resistivity in a conventional metallic state can be given by $R = m^*/(e^2 n\tau)$ (here $e$ is the elementary charge, $n$ is the carrier concentration and $\tau^{-1}$ is the scattering rate),[41] and (ii) the effective mass ratio $m_z^*/m_x^*$ is much less than the resistivity ratio $R_c/R_a$ at 10 K, the sudden increase in the electrical resistivity anisotropy of $WP_2$ at $T \leq 100$ K is unlikely to arise from the anisotropic effective mass of its electronic structure.

Optical spectroscopy is an efficient experimental technique for investigating charge dynamics in solids.[61-67] In order to further study the roles of the anisotropies of the carrier concentration $n$, effective mass $m*$ and scattering rate $\tau^{-1}$ in the emergence of the sharp increase in the electrical resistivity anisotropy, we carried out optical reflectance measurements on the $WP_2$ single crystals with ***E*** // $c$-axis and ***E*** // $a$-axis at different temperatures. **Figure** 4d displays several representative reflectance spectra $R(\omega)$. Sharp plasma edges are present in the low-energy regions of the $c$-axis and $a$-axis $R(\omega)$. The screened plasma frequencies of the $c$-axis and $a$-axis $R(\omega)$, which correspond to the energy positions of the minima of the plasma edges, $\omega_c^{src} \approx 2750$ cm$^{-1}$ and $\omega_a^{src} \approx 3651$ cm$^{-1}$ (see the two red arrows in **Figure** 4d). Given that (i) the square of the plasma frequency, which has the form $\omega_p^2 = 4\pi ne^2/m*$, is proportional to $n/m*$, and (ii) the $\omega_p^2$ scales linearly with the spectral weight (i.e., Drude weight $S$) of the Drude component of the real part of the optical conductivity $\sigma_1(\omega)$, which can be estimated by integrating the $\sigma_1(\omega)$ up to the screened plasma frequency, i.e., $\omega_p^2 = 8S \approx 8\int_0^{\omega^{src}} \sigma_1(\omega)d\omega$, we need to get the $\sigma_1(\omega)$ of the $WP_2$ single crystals by the Kramers-Kronig transformations of the measured $R(\omega)$ and then to obtain the spectral weight $S(\omega)$ of the $\sigma_1(\omega)$.[59-65] **Figure** 4e and **Figure** 4f show the $\sigma_1(\omega)$ and the $S(\omega)$, respectively (see the details about the Kramers-Kronig transformation in Experimental Section). Therein, the Drude weight $S$ corresponds to the $S(\omega)$ at the screened plasma frequency (see the red dashed lines in **Figure** 4f). As shown in **Figure** 4g, the ratio (i.e., $S_c^{-1}/S_a^{-1}$) between the reciprocals of the $c$-axis and $a$-axis Drude weights (i.e., $S_c^{-1}$ and $S_a^{-1}$) is about 1.3 and exhibits a quite weak dependence on $T$. Because (i) $(\omega_p^2)^{-1} \propto m*/n \propto S^{-1}$, (ii) the weak $T$-dependence of the $S_c^{-1}/S_a^{-1}$ is in sharp contrast to the abrupt increase in the $R_c/R_a$ at low temperatures, (iii) the $S_c^{-1}/S_a^{-1}$ is much smaller than the $R_c/R_a$ of ~ 8 at 10 K, the sudden increase in the electrical resistivity anisotropy of $WP_2$ should not be mainly caused by the anisotropy of the $m*/n$ (or the $S^{-1}$). Furthermore, we derived the scattering rate spectra $\tau^{-1}(\omega)$ by performing the extended Drude analysis of the optical data (see **Figure** 4h and the details about the extended Drude analysis in Experimental Section).[61-63] We plotted the *c*-axis and *a*-axis scattering rate at $\omega = 0$ (i.e., $\tau_c^{-1}$ and $\tau_a^{-1}$) as a function of temperature in the inset of **Figure** 4i. As shown in **Figure** 4i, the ratio $\tau_c^{-1}/\tau_a^{-1}$ between the $\tau_c^{-1}$ and $\tau_a^{-1}$ increases rapidly at $T \leq 100$ K and reaches the value of ~ 8.0 at 10 K, which is consistent with the *T*-dependence of the $R_c/R_a$. Therefore, the consistency between the *T*-dependences of the ratios $\tau_c^{-1}/\tau_a^{-1}$ and $R_c/R_a$ indicates that the rapid enhancement of the scattering rate anisotropy at $T \leq 100$ K is the main factor giving rise to the abrupt increase in the electrical resistivity anisotropy at low temperatures.

It is worth noticing that in contrast to the pairs of neighboring Weyl points with opposite chirality in the previously reported type-II WSM candidates, such as $MoTe_2$ and $TaIrTe_4$, the neighboring Weyl points in $WP_2$ have the same chirality.[32,38] For $WP_2$, this distinguished character—the protection of the neighboring Weyl points with the same chirality against annihilation upon structural distortions or defects can significantly reduce the probability of the scattering between its neighboring Weyl points.[38] Moreover, these neighboring Weyl points in $WP_2$ are located within the same $k_x$-$k_y$ plane, which implies that the scattering rate within the $k_x$-$k_y$ plane may be much smaller than the scattering rates within the other planes including the $k_z$ direction. Therefore, one possible physical source of the scattering rate anisotropy (i.e., $\tau_c^{-1}/\tau_a^{-1}$) may be related to the significantly reduced probability of the scattering between its neighboring Weyl points with the same chirality in the same $k_x$-$k_y$ plane. In this regard, since the pairs of neighboring Weyl points in the reported type-II WSM candidates have opposite chirality, the neighboring Weyl points with the same chirality in the same $k_x$-$k_y$ plane may make the scattering rate anisotropy in $WP_2$ different from those in the reported type-II WSM candidates, such as $MoTe_2$ and $TaIrTe_4$. In addition, previous theoretical studies suggest that for a WSM at low temperatures, its disorders and the anisotropy of the Fermi velocity of Weyl fermions can induce the resistivity anisotropy, e.g., $R_c/R_a \sim (\upsilon_a/\upsilon_c)^2$, where $\upsilon_a$ and $\upsilon_c$ are the Fermi velocities of Weyl fermions along the *a*-axis and *c*-axis, respectively.[68] In $WP_2$, $\upsilon_a/\upsilon_c \sim 10$,[32] so $R_c/R_a \sim 100$, which is much larger than the measured resistivity anisotropy at $T = 10$ K. Perhaps, its resistivity anisotropy induced by the disorders and the Fermi velocity anisotropy may be weakened by its topologically trivial bands and then reach the anisotropy degree of its resistivity at $T = 10$ K.

To further study whether the electrical resistivity ratio $R_c/R_a$ of the $WP_2$ single crystals can be tuned, we performed the electrical resistivity measurements at $T = 10$ K with the magnetic field applied perpendicular to the crystalline *ac*-plane. **Figure** 5a displays the magnetic-field dependent resistivities measured with the electric currents along the *c*-axis and *a*-axis, respectively. It can be observed that the difference between the $R_c$ and the $R_a$ grows continuously with increasing magnetic fields. Then, we plotted the electrical resistivity ratio $R_c(10\text{ K})/R_a(10\text{ K})$ at $T = 10$ K as a function of magnetic field ($B$) in **Figure** 5b. The electrical resistivity ratio $R_c(10\text{ K})/R_a(10\text{ K})$ increases from ~ 8.0 to ~ 10.6 as the magnetic field is enhanced from 0 to 9 T, which manifests that the magnetic field can effectively tune the resistivity ratio $R_c(10\text{ K})/R_a(10\text{ K})$ (see **Figure** 5c). The strong and tunable anisotropy of the in-plane (i.e., *ac*-plane) electrical resistivity found in $WP_2$ can act as a degree of freedom for tuning the electrical properties of type-II WSMs and thus have potential applications in new generation of electronic devices.[69-71]

In summary, we have performed the electrical transport, optical spectroscopy, first-principle and ARPRS studies of the type-II WSM candidate in $WP_2$. The ratio ($R_c/R_a$) between the crystalline *c*-axis and *a*-axis electrical resistivities shows a weak dependence on temperature in the temperature range from 100 to 250 K, but increases suddenly at $T \leq 100$ K, and then reaches the value of ~ 8.0 at $T = 10$ K, which is by far the strongest in-plane electrical resistivity anisotropy among the known type-II WSM candidates with their carrier concentrations comparable to ~ $10^{21}$ $cm^{-3}$. Our optical spectroscopy study, combined with the first-principle calculations on the electronic band structure, indicates that the sudden increase in the electrical resistivity anisotropy at $T \leq 100$ K mainly originates from the sharp enhancement of the scattering rate anisotropy at low temperatures. Moreover, the $R_c/R_a$ of $WP_2$ at $T = 10$ K can be tuned from 8.0 to 10.6 as the magnetic field was enhanced from 0 to 9

T. The strong and tunable electrical resistivity anisotropy found in $WP_2$ can be considered as a degree of freedom for tuning the electrical properties of type-II WSMs. In addition, the good agreement between the angle dependences of the Raman-active phonon peak intensities measured using ARPRS and the fitting results based on the broken spatial inversion symmetry of the $WP_2$ crystal structure, together with the consistence between the energies of the peaks observed in the polarized Raman spectra and the calculated Raman-active phonon energies, provides convincing evidence for the spatial inversion symmetry breaking in the type-II WSM candidate $WP_2$. Our work not only opens a new avenue for the experimental identification of the type-II WSM state in $WP_2$ but also paves the way for developing a new generation of electronic devices based on the type-II WSMs with strong and tunable electrical anisotropy.

## Experimental Section

*Growth and characterization of the $WP_2$ single crystals*: The single crystals of $WP_2$ investigated in this paper were prepared via chemical vapor transport (CVT) method.[47] The $WO_3$ (Sigma Aldrich, 99:995%), red phosphorus (Alfa Aesar, 99:999%) and iodine (Alfa Aesar, 99:99%) were sealed in an evacuated quartz tube. The quartz tube was placed in a tube furnace and gradually heated up to a temperature gradient of 970 ℃ - 850 ℃. The reaction was held under this temperature gradient for 12 days. Then, the needle like $WP_2$ single crystals were obtained. The crystal structure of the as-grown single crystals was analyzed by single crystal X-ray diffractometer (SXRD, BRUKER D8 VENTURE) with Mo-Kα radiation (0.7 Å). The determined lattice parameters $a$ = 3.17 Å, $b$ = 11.18 Å and $c$ = 4.98 Å, which are the same as the previous experimental studies.[72,73] **Figure** S1 shows the indices of crystallographic plane of the $WP_2$ single crystals.

*Angle-resolved polarized Raman measurements:* The angle-resolved polarized Raman measurements were performed in a backscattering geometry using a HORIBA LabRAM HR Evolution Raman spectrometer. The rotation angle of each step for the crystal-angle-resolved Raman experiments is 10°.

*First-principle calculations:* The first-principle calculations were performed using the Vienna *ab initio* simulation package (VASP)[74] with the generalized gradient approximation (GGA) of the Perdew-Burke-Ernzerhof functional.[75,76] The plane-wave kinetic energy cutoff of 520 eV and the 9 × 9 × 9 Γ-centered k-point mesh were set. The atomic coordinates and the lattice shapes are fully relaxed until the Hellmann–Feynman forces on the atoms all are less than 0.01 eV Å$^{-1}$ and the convergence criteria for energy was set at $10^{-8}$ eV. The initial lattice parameters of $a$ = 3.181 Å, $b$ = 11.237 Å, and $c$ = 5.007 Å were used.[77] To determine the phonon vibrational modes and phonon frequencies of $WP_2$ at the Brillouin zone center, the finite displacement method implemented in phonopy package was used.[78] The electronic band structure of $WP_2$ shown in **Figure** 4c of the main text was calculated with considering spin-orbit coupling.

*Electrical resistivity measurements:* The electrical resistivity measurements were carried out in the low temperature superconducting magnet thermostat system (TKMS). The electrical resistivities were measured using a standard four-probe method with the electric currents applied along the *a*-axis and *c*-axis, respectively.

*Optical reflectance measurements:* The optical reflectance measurements were carried out on a Bruker Vertex 80v spectrometer in the energy range up to 25000 cm$^{-1}$. The single-crystal samples were mounted on the optically black cone locating at the cold finger of a helium flow cryostat. An *in situ* gold and aluminum overcoating technique was employed to get the optical reflectance spectra $R(\omega)$. Linear polarizers were used for obtaining the linearly polarized incident light with its electric field of the incident light ***E*** // *c*-axis and ***E*** // *a*-axis. The optical reflectance data are highly reproducible.

*Kramers-Kronig transformation:* A Kramers-Kronig transformation of the $R(\omega)$ of the $WP_2$ single crystals was used for getting the phase shift (i.e., $\theta(\omega)$) of the reflected light relative to the incident light according to the following relationship:

$$\theta(\omega) = -\frac{\omega}{\pi} P \int_0^{+\infty} \frac{\ln R(\omega')}{{\omega'}^2 - \omega^2} d\omega' \quad (4)$$

where $P$ denotes the Cauchy principal value. A Hagen-Rubens relation was used for low-energy extrapolation, and a $\omega^{-0.3}$ dependence was used for the high-energy extrapolation up to 300000 cm$^{-1}$, above which a $\omega^{-4}$ dependence is employed. The $R(\omega)$ and $\theta(\omega)$ have the relationships with the real part $n(\omega)$ and the imaginary part $k(\omega)$ of the refractive index:

$$n(\omega) = \frac{1-R(\omega)}{1+R(\omega)-2\sqrt{R(\omega)}cos(\theta(\omega))} \quad (5)$$

$$k(\omega) = \frac{-2\sqrt{R(\omega)}\sin(\theta(\omega))}{1+R(\omega)-2\sqrt{R(\omega)}cos(\theta(\omega))} \quad (6)$$

Moreover, the real part $\sigma_1(\omega)$ and the imaginary part $\sigma_2(\omega)$ of the optical conductivity have the following relationships with $n(\omega)$ and $k(\omega)$:

$$\sigma_1(\omega) = \frac{\omega n(\omega)k(\omega)}{2\pi} \quad (7)$$

$$\sigma_2(\omega) = \frac{\omega\left(1-n^2(\omega)+k^2(\omega)\right)}{4\pi} \quad (8)$$

Therefore, the real part $\sigma_1(\omega)$ and the imaginary part $\sigma_2(\omega)$ of the optical conductivity can be obtained when we have gotten the $R(\omega)$ and $\theta(\omega)$.

*Extended Drude analysis of the optical data:* The scattering rate spectra $\tau^{-1}(\omega)$ can be obtained by performing the extended Drude analysis of the optical data according to Equation (9) in Ref. [61]:

$$\tau^{-1}(\omega) = \frac{\omega_p^2}{4\pi}Re\left(\frac{1}{\sigma(\omega)}\right) = \frac{\omega_p^2}{4\pi}\frac{\sigma_1(\omega)}{\sigma_1^2(\omega)+\sigma_2^2(\omega)} \quad (9)$$

where $\omega_p$ is the plasma frequency, $\sigma(\omega)$ is the complex optical conductivity, $Re$ denotes the real part of the $1/\sigma(\omega)$, $\sigma_1(\omega)$ is the real part of the $\sigma(\omega)$, and $\sigma_2(\omega)$ is the imaginary part of the $\sigma(\omega)$. In the plasma frequency, the effective mass is an averaged mass of the charge carriers occupying the bands at Fermi energy.

## Supporting Information

Supporting Information is available from the Wiley Online Library or from the author.

## Acknowledgements

†B.S., Y.S. and Y.H. contributed equally to this work. B.S. and Y.H. grew the single crystals. B.S., Y.H. and Y.Y. carried out the Raman measurements. B.S. did first-principle calculations and carried out the optical measurements. Y.S., X.C., B.S. and J.G. performed the transport experiments. B.S., J.Z., Y.M., J.L. and Z.-G.C. analyzed the data. B.S. and Z.-G.C. wrote the paper. Z.-G.C. conceived and supervised this project. We thank Rico U. Schoenemann, Fanming Qu, Di Chen, Meng Lv, Guang Yang, Nan Xu and Quansheng Wu for their help and discussions. The authors acknowledge support from the National Key Research and Development Program of China (Projects No. 2017YFA0304700, and No. 2016YFA0300600) and the Pioneer Hundred Talents Program of the Chinese Academy of Sciences.

## Conflict of Interest

The authors declare no conflict of interest.

## References

[1] X. Wan, A. M. Turner, A. Vishwanath, S. Y. Savrasov, *Phys. Rev. B* **2011,** 83, 205101.

[2] A. A. Burkov, L. Balents, *Phys. Rev. Lett.* **2011,** 107, 127205.

[3] H. Weng, C. Fang, Z. Fang, B. A. Bernevig, X. Dai, *Phys. Rev. X* **2015,** 5, 011029.

[4] S.-M. Huang, S.-Y. Xu, I. Belopolski, C.-C. Lee, G. Chang, B. Wang, N. Alidoust, G. Bian, M. Neupane, C. Zhang, S. Jia, A. Bansil, H. Lin, M. Z. Hasan, *Nat. Commun.* **2015,** 6, 7373.

[5] S.-Y. Xu, I. Belopolski, N. Alidoust, M. Neupane, G. Bian, C. Zhang, R. Sankar, G. Chang, Z. Yuan, C.-C. Lee, S.-M. Huang, H. Zheng, J. Ma, D. S. Sanchez, B. Wang, A. Bansil, F. Chou, P. P. Shibayev, H. Lin, S. Jia, M. Z. Hasan, *Science* **2015,** 349, 613.

[6] C. Shekhar, A. K. Nayak, Y. Sun, M. Schmidt, M. Nicklas, I. Leermakers, U. Zeitler, Y. Skourski, J. Wosnitza, Z. Liu, Y. Chen, W. Schnelle, H. Borrmann, Y. Grin, C. Felser, B. Yan, *Nat. Phys.* **2015,** 11, 645.

[7] L. X. Yang, Z. K. Liu, Y. Sun, H. Peng, H. F. Yang, T. Zhang, B. Zhou, Y. Zhang, Y. F. Guo, M. Rahn, D. Prabhakaran, Z. Hussain, S.-K. Mo, C. Felser, B. Yan, Y. L. Chen, *Nat. Phys.* **2015,** 11, 728.

[8] Z. K. Liu, L. X. Yang, Y. Sun, T. Zhang, H. Peng, H. F. Yang, C. Chen, Y. Zhang, Y. F. Guo, D. Prabhakaran, M. Schmidt, Z. Hussain, S.-K. Mo, C. Felser, B. Yan, Y. L. Chen, *Nat. Mater.* **2015,** 15, 27.

[9] B. Q. Lv, N. Xu, H. M.Weng, J. Z. Ma, P. Richard, X. C. Huang, L. X. Zhao, G. F. Chen, C. E. Matt, F. Bisti, V. N. Strocov, J. Mesot, Z. Fang, X. Dai, T. Qian, M. Shi, H. Ding, *Nat. Phys.* **2015**, 11, 724.

[10] S.-Y. Xu, N. Alidoust, I. Belopolski, Z. Yuan, G. Bian, T.-R. Chang, H. Zheng, V. N. Strocov, D. S. Sanchez, G. Chang, C. Zhang, D. Mou, Y. Wu, L. Huang, C.-C. Lee, S.-M. Huang, B. Wang, A. Bansil, H.-T. Jeng, T. Neupert, A. Kaminski, H. Lin, S. Jia, M. Z. Hasan, *Nat. Phys.* **2015,** 11, 748.

[11] A. A. Zyuzin, A. A. Burkov, *Phys. Rev. B* **2012,** 86, 115133.

[12] M. M. Vazifeh, M. Franz, *Phys. Rev. Lett.* **2013,** 111, 027201.

[13] C.-X. Liu, P. Ye, X.-L. Qi, *Phys. Rev. B* **2013,** 87, 235306.

[14] X. Huang, L. Zhao, Y. Long, P. Wang, D. Chen, Z. Yang, H. Liang, M. Xue, H. Weng, Z. Fang, X. Dai, G. Chen, *Phys. Rev. X* **2015,** 5, 031023.

[15] C.-L. Zhang, S.-Y. Xu, I. Belopolski, Z. Yuan, Z. Lin, B. Tong, G. Bian, N. Alidoust, C.-C. Lee, S.-M. Huang, T.-R. Chang, G. Chang, C.-H. Hsu, H.-T. Jeng, M. Neupane, D. S. Sanchez, H. Zheng, J. Wang, H. Lin, C. Zhang, H.-Z. Lu, S.-Q. Shen, T. Neupert, M. Z. Hasan, S. Jia, *Nat. Commun.* **2016,** 7, 10735.

[16] M. Hirschberger, S. Kushwaha, Z. Wang, Q. Gibson, S. Liang, C. A. Belvin, B. A. Bernevig, R. J. Cava, N. P. Ong, *Nat. Mater.* **2016,** 15, 1161.

[17] K. Kuroda, T. Tomita, M.-T. Suzuki, C. Bareille, A. A. Nugroho, P. Goswami, M. Ochi, M. Ikhlas, M. Nakayama, S. Akebi, R. Noguchi, R. Ishii, N. Inami, K. Ono, H. Kumigashira, A. Varykhalov, T. Muro, T. Koretsune, R. Arita, S. Shin, Takeshi Kondo, S. Nakatsuji, *Nat. Mater.* **2017,** 16, 1090.

[18] C. R. Rajamathi, U. Gupta, N. Kumar, H. Yang, Y. Sun, V. Süß, C Shekhar, M. Schmidt, H. Blumtritt, P. Werner, B. Yan, S. Parkin, C. Felser, C. N. R. Rao, *Adv. Mater.* **2017,** 29, 1606202.

[19] A. Politano, G. Chiarello, Z. Li, V. Fabio, L. Wang, L. Guo, X. Chen, D. Boukhvalov, *Adv. Funct. Mater.* **2018,** 28, 1800511.

[20] G. Xu, H. Weng, Z. Wang, X. Dai, Z. Fang, *Phys. Rev. Lett.* **2011,** 107, 186806.

[21] J. Ruan, S.-K. Jian, D. Zhang, H. Yao, H. Zhang, S.-C. Zhang, D. Xing, *Phys. Rev. Lett.* **2016,** 116, 226801.

[22] A. A. Zyuzin, S. Wu, A. A. Burkov, *Phys. Rev. B* **2012,** 85, 165110.

[23] Z. Wang, M. G. Vergniory, S. Kushwaha, M. Hirschberger, E. V. Chulkov, A. Ernst, N. P. Ong, R. J. Cava, B. A. Bernevig, *Phys. Rev. Lett.* **2016,** 117, 236401.

[24] H. Yang, Y. Sun, Y. Zhang, W.-J. Shi, S. S P Parkin, B. Yan, *New J. Phys.* **2017,** 19 015008.

[25] J. Y. Liu, J. Hu, Q. Zhang, D. Graf, H. B. Cao, S. M. A. Radmanesh, D. J. Adams, Y. L. Zhu, G. F. Cheng, X. Liu, W. A. Phelan, J. Wei, M. Jaime, F. Balakirev, D. A. Tennant, J. F. DiTusa, I. Chiorescu, L. Spinu, Z. Q. Mao, *Nat. Mater.* **2017,** 16, 905.

[26] S. Huang, J. Kim, W. A. Shelton, E. W. Plummer, R. Jin, *Proc. Natl. Acad. Sci. USA* **2017,** 114, 6256.

[27] C. Guo, C. Cao, M. Smidman, F. Wu, Y. Zhang, F. Steglich, F.-C. Zhang, H. Yuan, *npj Quantum Mater.* **2017,** 2, 39.

[28] A. A. Soluyanov, D. Gresch, Z. Wang, Q. Wu, M. Troyer, X. Dai, B. A. Bernevig, *Nature* **2015,** 527, 495.

[29] Y. Sun, S.-C. Wu, M. N. Ali, C. Felser, B. Yan, *Phys. Rev. B* **2015,** 92, 161107(R).

[30] Z. Wang, D. Gresch, A. A. Soluyanov, W. Xie, S. Kushwaha, X. Dai, M. Troyer, R. J. Cava, B. A. Bernevig, *Phys. Rev. Lett.* **2016,** 117, 056805.

[31] G. Chang, S.-Y. Xu, D. S. Sanchez, S.-M. Huang, C.-C. Lee, T.-R. Chang, G. Bian, H. Zheng, I. Belopolski, N. Alidoust, H.-T. Jeng, A. Bansil, H. Lin, M. Z. Hasan, *Sci. Adv.* **2016,** 2, e1600295.

[32] G. Autès, D. Gresch, M. Troyer, A. A. Soluyanov, O. V. Yazyev, *Phys. Rev. Lett.* **2016,** 117, 066402.

[33] S.-Y. Xu, N. Alidoust, G. Chang, H. Lu, B. Singh, I. Belopolski, D. S. Sanchez, X. Zhang, G. Bian, H. Zheng, M.-A. Husanu, Y. Bian, S.-M. Huang, C.-H. Hsu, T.-R. Chang,

H.-T. Jeng, A. Bansil, T. Neupert, V. N. Strocov, H. Lin, S. Jia, M. Z. Hasan, *Sci. Adv.* **2017,** 3, e1603266.

[34] Y.-Y. Lv, X. Li, B.-B. Zhang, W. Y. Deng, S.-H. Yao, Y. B. Chen, J. Zhou, S.-T. Zhang, M.-H. Lu, L. Zhang, M. Tian, L. Sheng, Y.-F. Chen, *Phys. Rev. Lett.* **2017,** 118, 096603.

[35] Z.-M. Yu, Y. Yao, S. A. Yang, *Phys. Rev. Lett.* **2016,** 117, 077202.

[36] M. Udagawa, E. J. Bergholtz, *Phys. Rev. Lett.* **2016,** 117, 086401.

[37] S. Tchoumakov, M. Civelli, M. O. Goerbig, *Phys. Rev. Lett.* **2016,** 117, 086402.

[38] N. Kumar, Y. Sun, N. Xu, K. Manna, M. Yao, V. Süss, I. Leermakers, O. Young, T. Förster, M. Schmidt, H. Borrmann, B. Yan, U. Zeitler, M. Shi, C. Felser, C. Shekhar, *Nature Commun.* **2017,** 8, 1642.

[39] A. Wang, D. Graf, Y. Liu, Q. Du, J. Zheng, H. Lei, C. Petrovic, *Phys. Rev. B* **2017,** 96, 121107(R).

[40] R. Schönemann, N. Aryal, Q. Zhou, Y.-C. Chiu, K.-W. Chen, T. J. Martin, G. T. McCandless, J. Y. Chan, E. Manousakis, L. Balicas, *Phys. Rev. B* **2017,** *96,* 121108(R).

[41] J. Gooth, F. Menges, N. Kumar, V. Süβ, C. Shekhar, Y. Sun, U. Drechsler, R. Zierold, C. Felser, B. Gotsmann, *Nature Commun.* **2018,** 9, 4093.

[42] A. Jaoui, B. Fauqué, C. W. Rischau, A. Subedi, C. Fu, J. Gooth, N. Kumar, V. Süß, D. L. Maslov, C. Felser, K. Behnia, *npj Quan. Mater.* **2018,** 3, 64.

[43] J. Coulter, R. Sundararaman, P. Narang, *Phys. Rev. B* **2018,** 98, 115130.

[44] E. Razzoli, B. Zwartsenberg, M. Michiardi, F. Boschini, R. P. Day, I. S. Elfimov, J. D. Denlinger, V. Süss, C. Felser, A. Damascelli, *Phys. Rev. B* **2018,** 97, 201103(R).

[45] M.-Y. Yao, N. Xu, Q. S. Wu, G. Autès, N. Kumar, V. N. Strocov, N. C. Plumb, M. Radovic, O. V. Yazyev, C. Felser, J. Mesot, M. Shi, *Phys. Rev. Lett.* **2019,** 122, 176402.

[46] R. Loudon, *Adv. Phys.* **1964,** 13, 423.

[47] R. Rühl, W. Jeitschko, *Monatsh. Chem.* **1983,** 114, 817-828.

[48] J. H. Du, Z. F. Lou, S. N. Zhang, Y. X. Zhou, B. J. Xu, Q. Chen, Y. Q. Tang, S. J. Chen, H. C. Chen, Q. Q. Zhu, H. D. Wang, J. H. Yang, Q. S. Wu, V. Oleg, Yazyev, M. H. Fang, *Phys. Rev. B* **2018,** 97, 245101.

[49] L. Zhou, S. Huang, Y. Tatsumi, L. Wu, H. Guo, Y.-Q. Bie, K. Ueno, T. Yang, Y. Zhu, J. Kong, R. Saito, M. Dresselhaus, *J. Am. Chem. Soc.* **2017,** 139, 8396.

[50] Y. Liu, Q. Gu, Y. Peng, S. Qi, N. Zhang, Y. Zhang, X. Ma, R. Zhu, L. Tong, J. Feng, Z. Liu, J.-H. Chen, *Adv. Mater.* **2018,** 30, 1706402.

[51] Q. Song, X. Pan, H. Wang, K. Zhang, Q. Tan, P. Li, Y. Wan, Y. Wang, X. Xu, M. Lin, X. Wan, F. Song, L. Dai, *Sci. Rep.* **2016,** 6, 29254.

[52] M. Kim, S. Han, J. H. Kim, J.-U. Lee, Z. Lee, H. Cheong, *2D Mater.* **2016,** 3, 034004.

[53] S. Tongay, H. Sahin, C. Ko, A. Luce, W. Fan, K. Liu, J. Zhou, Y.-S. Huang, C.-H. Ho, J. Yan, D. F. Ogletree, S. Aloni, J. Ji, S. Li, J. Li, F. M. Peeters, J. Wu, *Nat. Commun.* **2014**, 5, 3252.

[54] R. He, J.-A. Yan, Z. Yin, Z. Ye, G. Ye, J. Cheng, J. Li, C. H. Lui, *Nano lett.* **2016,** 16, 1404.

[55] K. Zhang, C. Bao, Q. Gu, X. Ren, H. Zhang, K. Deng, Y. Wu, Y. Li, J. Feng, S. Zhou, *Nature Commun.* **2016,** 7, 13552.

[56] S. Y. Chen, T. Goldstein, D. Venkataraman, et al. *Nano Lett.* **2016,** 16, 5852.

[57] R. Beams, L. G. Cançado, S. Krylyuk, I. Kalish, B. Kalanyan, A. K. Singh, K. Choudhary, A. Bruma, P. M. Vora, F. Tavazza, A. V. Davydov, S. J. Stranick, *ACS Nano* **2016,** 10, 9626.

[58] J. Wang, X. Luo, S. Li, I. Verzhbitskiy, W. Zhao, S. Wang, S. Y. Quek, G. Eda, *Adv. Funct. Mater.* **2017,** 27, 1604799.

[59] H. P. Hughes, R. H. Friend, *J. Phys. C: Solid State Phys.* **1978,** 11, L103.

[60] D. H. Keum, S. Cho, J. H. Kim, D.-H. Choe, H.-J. Sung, M. Kan, H. Kang, J.-Y. Hwang, S. W. Kim, H. Yang, K. J. Chang, Y. H. Lee, *Nature Phys.* **2015,** 11, 482.

[61] D. N. Basov, T. Timusk, *Rev. Mod. Phys.* **2005,** 77, 721.

[62] D. N. Basov, R. D. Averitt, D. van der Marel, M. Dressel, K. Haule, *Rev. Mod. Phys.* **2011,** 83, 471.

[63] M. Dressel, G. Grüner, *Electrodynamics of Solids: Optical Properties of Electrons in Matter*, Cambridge University Press, Cambridge **2002**.

[64] S. Y. Kim, M.-C. Lee, G. Han, M. Kratochvilova, S. Yun, S. J. Moon, C. Sohn, J.-G. Park, C. Kim, T. W. Noh, *Adv. Mater.* **2018,** 30, 1704777.

[65] Z. G. Chen, T. Dong, R. H. Ruan, B. F. Hu, B. Cheng, W. Z. Hu, P. Zheng, Z. Fang, X. Dai, N. L. Wang, *Phys. Rev. Lett.* **2010,** 105, 097003.

[66] Y. M. Dai, B. Xu, B. Shen, H. Xiao, H. H. Wen, X. G. Qiu, C. C. Homes, R. P. S. M. Lobo, *Phys. Rev. Lett.* **2013,** 111, 117001.

[67] B. Xu, L. X. Zhao, P. Marsik, E. Sheveleva, F. Lyzwa, Y. M. Dai, G. F. Chen, X. G. Qiu, C. Bernhard, *Phys. Rev. Lett.* **2018,** 121, 187401.

[68] Ya. I. Rodionov, K. I. Kugel, N. Franco, *Phys. Rev. B* **2015,** 92, 195117.

[69] Q. Ma, S.Y. Xu, C. K. Chan, C. L. Zhang, G. Q. Chang, Y. X. Lin, W. W. Xie, T. Palacios, H. Lin, S. Jia, A. L. Patrick, J. H. Pablo, G, Nuh, *Nat. Phys.* **2017,** 13, 842−847.

[70] D. Sun, Z. K. Wu, D. Charles, X. B. Li, B. Claire, A. D. H. Walt, N. F. Phillip, B. N. Theodore, *Phys. Rev. Lett.* **2008,** 101, 157402.

[71] J. W. Lai, Y. N. Liu, J. C. Ma, X. Zhuo, Y. Peng, W. Lu, Z. Liu, J. H. Chen, D. Sun, *ACS nano* **2018,** 12, 4055-4061.

[72] H. Mathis, R. Glaum, R. Gruehn, *Acta. Chem. Scand.* **1991,** 45, 781.

[73] S. Rundqvist, T. Lundstrom, *Acta Chem. Scand.* **1963,** 17, 37.

[74] G. Kresse, J. Furthmüller, *Phys. Rev. B* **1996,** 54, 11169.

[75] J. P. Perdew, J. A. Chevary, S. H. Vosko, K. A. Jackson, M. R. Pederson, D. J. Singh, C. Fiolhais, *Phys. Rev. B* **1992,** 46, 6671.

[76] J. P. Perdew, K. Burke, M. Ernzerhof, *Phys. Rev. Lett.* **1996,** 77, 3865.

[77] A. Jain, S. P. Ong, G. Hautier, W. Chen, W. D. Richards, S. Dacek, S. Cholia, D. Gunter, D. Skinner, G. Cederand, K. A. Persson, *Apl Mater.* **2013,** 1, 011002.

[78] X. Gonze, C. Lee, *Phys. Rev. B* **1997,** 55, 10355.

## Figures, Figure Legends and Table

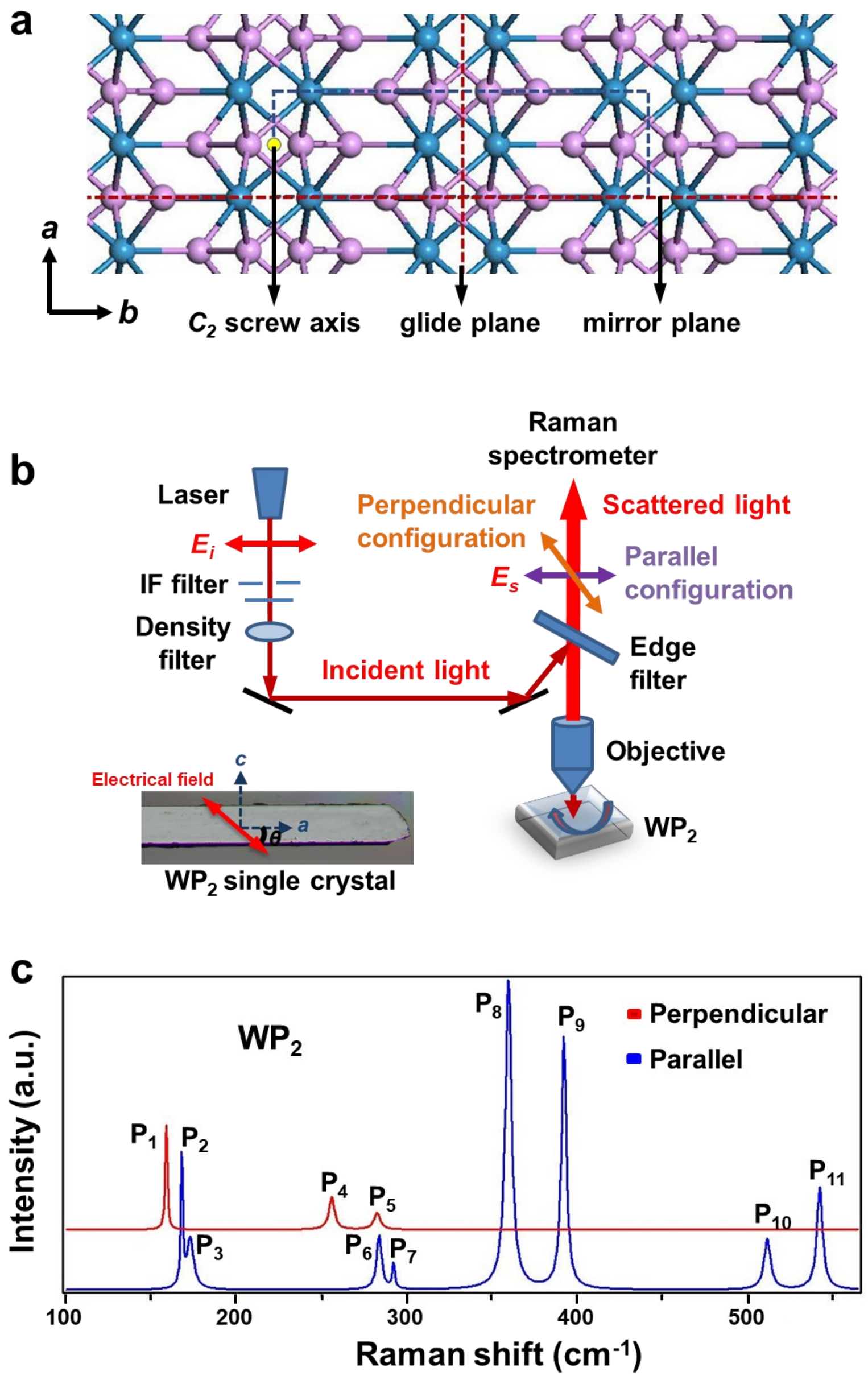

**Figure 1.** Crystal structure and polarized Raman spectra of $WP_2$. a) Crystal structure of $WP_2$. The W and P atoms are denoted by the light cyan and pink spheres, respectively. The cyan dashed box shows the orthorhombic conventional unit cell. The $C_2$ screw axis along the crystalline $c$-axis is shown by the yellow dot. The glide and mirror planes are indicated by the two red dashed lines, respectively. b) Schematic of the angle-resolved polarized Raman spectroscopy (ARPRS). The electrical fields of the linearly polarized incident and scattered lights are $\boldsymbol{E_i}$ and $\boldsymbol{E_s}$, respectively. The bottom-left inset depicts the optical image of the $WP_2$ single crystal measured by ARPRS. c) Raman spectra of the $WP_2$ single crystal which were measured in the parallel-polarized (i.e., $\boldsymbol{E_i}$ // $\boldsymbol{E_s}$ // $a$-axis) and perpendicular-polarized (i.e., $\boldsymbol{E_i} \perp \boldsymbol{E_s}$, $\boldsymbol{E_i}$ // $a$-axis, and $\boldsymbol{E_s}$ // $c$-axis) configurations, respectively. Several peak-like features are present in the polarized Raman spectra of $WP_2$.

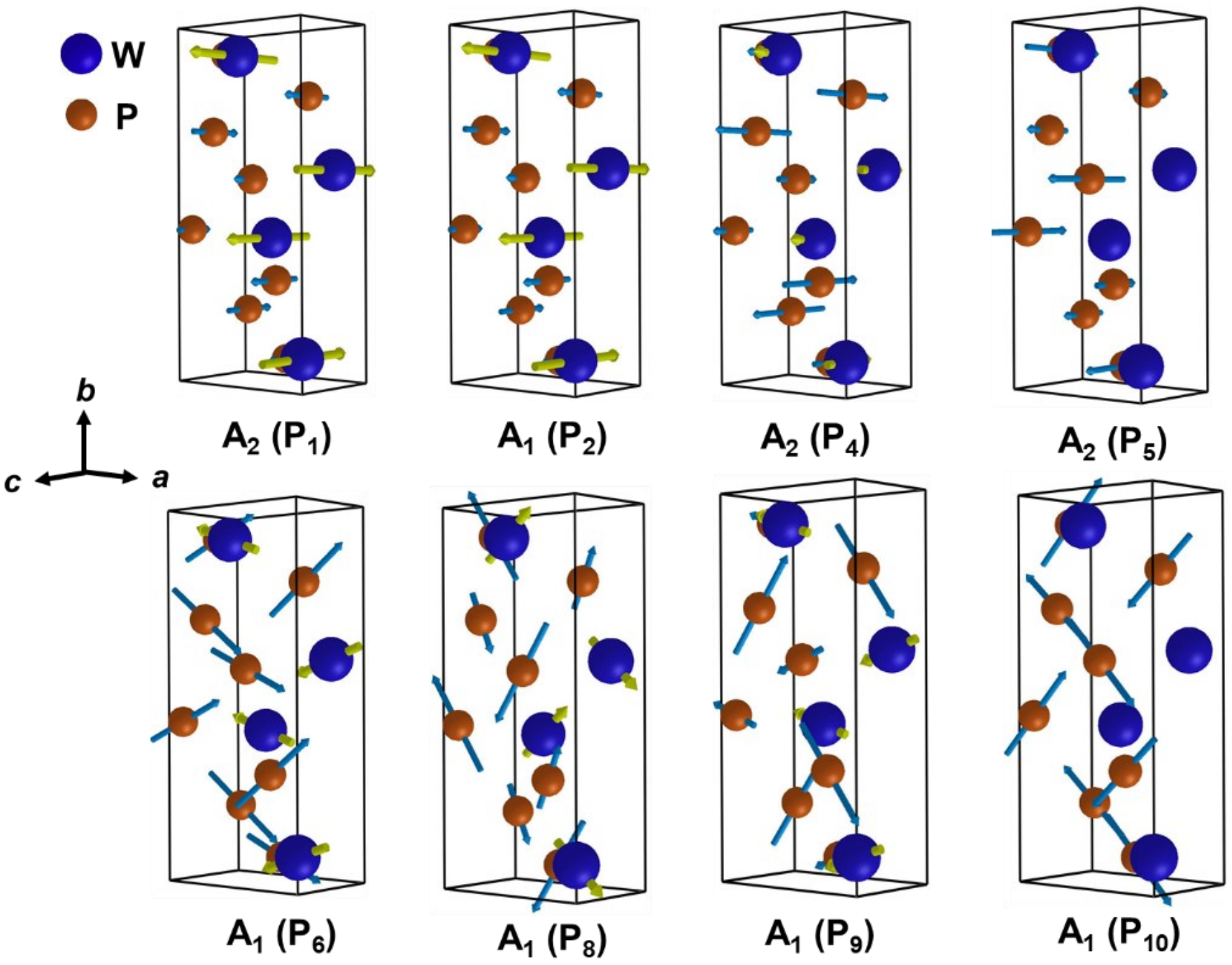

**Figure 2.** Atomic schematics of the calculated vibrational patterns of the Raman-active phonon modes $A_1$ and $A_2$. The Raman-active modes $A_1$ and $A_2$ of $WP_2$ can be observed under the parallel-polarized (i.e., $\boldsymbol{E_i}$ // $\boldsymbol{E_s}$ // *a*-axis) and perpendicular-polarized (i.e., $\boldsymbol{E_i} \perp \boldsymbol{E_s}$, $\boldsymbol{E_i}$ // *a*-axis and $\boldsymbol{E_s}$ // *c*-axis) configurations, respectively. The three measured peak-like features, $P_1$, $P_4$, and $P_5$, can be assigned to the three Raman-active phonon modes $3A_2$. The five measured peak-like features, $P_2$, $P_6$, $P_8$, $P_9$, and $P_{10}$ can be ascribed to the five Raman-active modes $5A_1$. The W and P atoms in the unit cell are denoted by blue and brown spheres, respectively. The atomic displacements of the W and P atoms are indicated by the colored arrows.

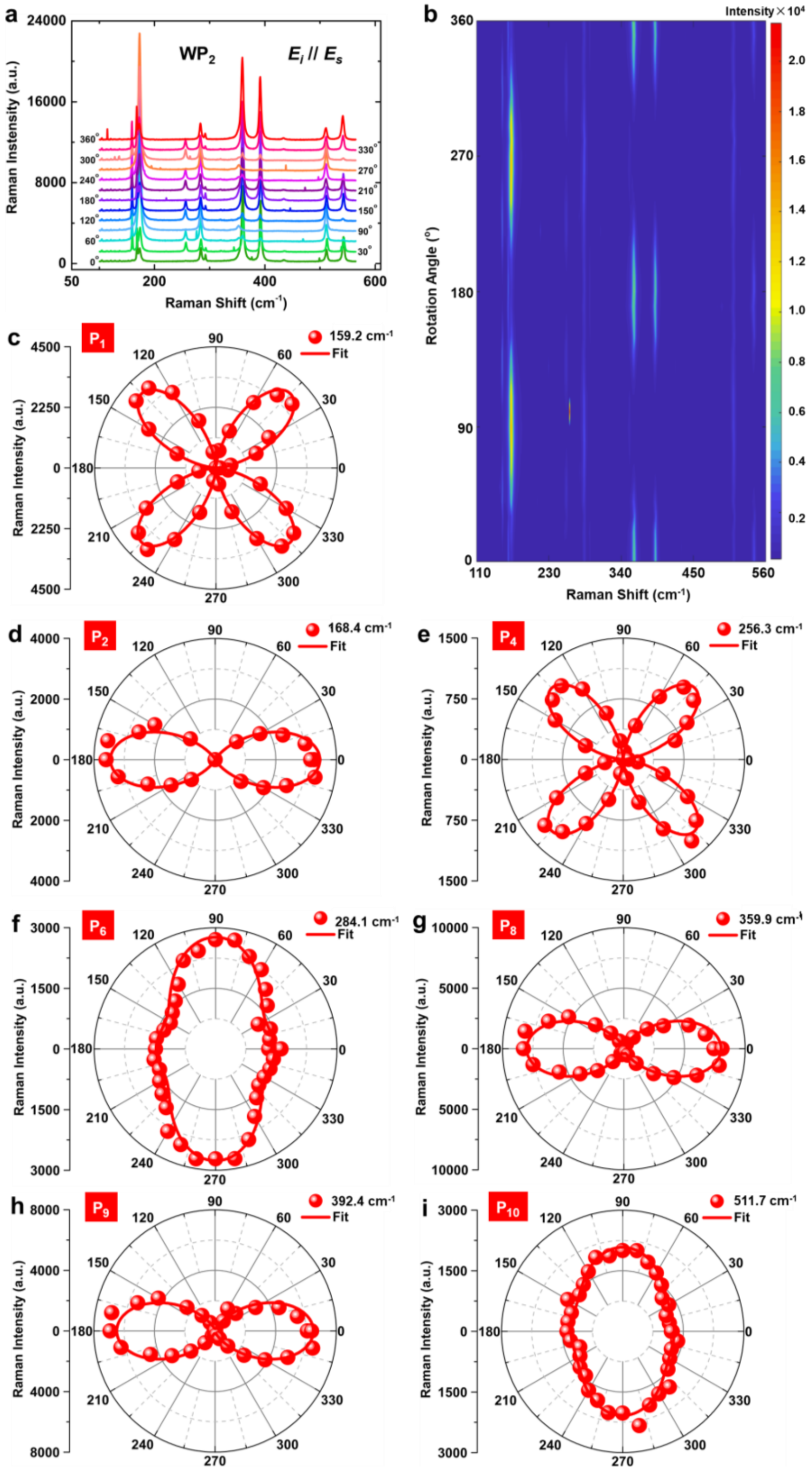

**Figure 3.** Angle dependences of the intensities of the $A_1$ and $A_2$ phonon peaks measured in the parallel-polarized configuration. a) Angle-dependent Raman spectra of the measured $WP_2$ single crystals. b) Contour map of the measured Raman phonon intensities as a function of angle and energy. c-i) Angle dependences of the Raman-active phonon peak intensities. The red dots show the measured data. The red solid curves display the fitting results based on Equations (2) and (3).

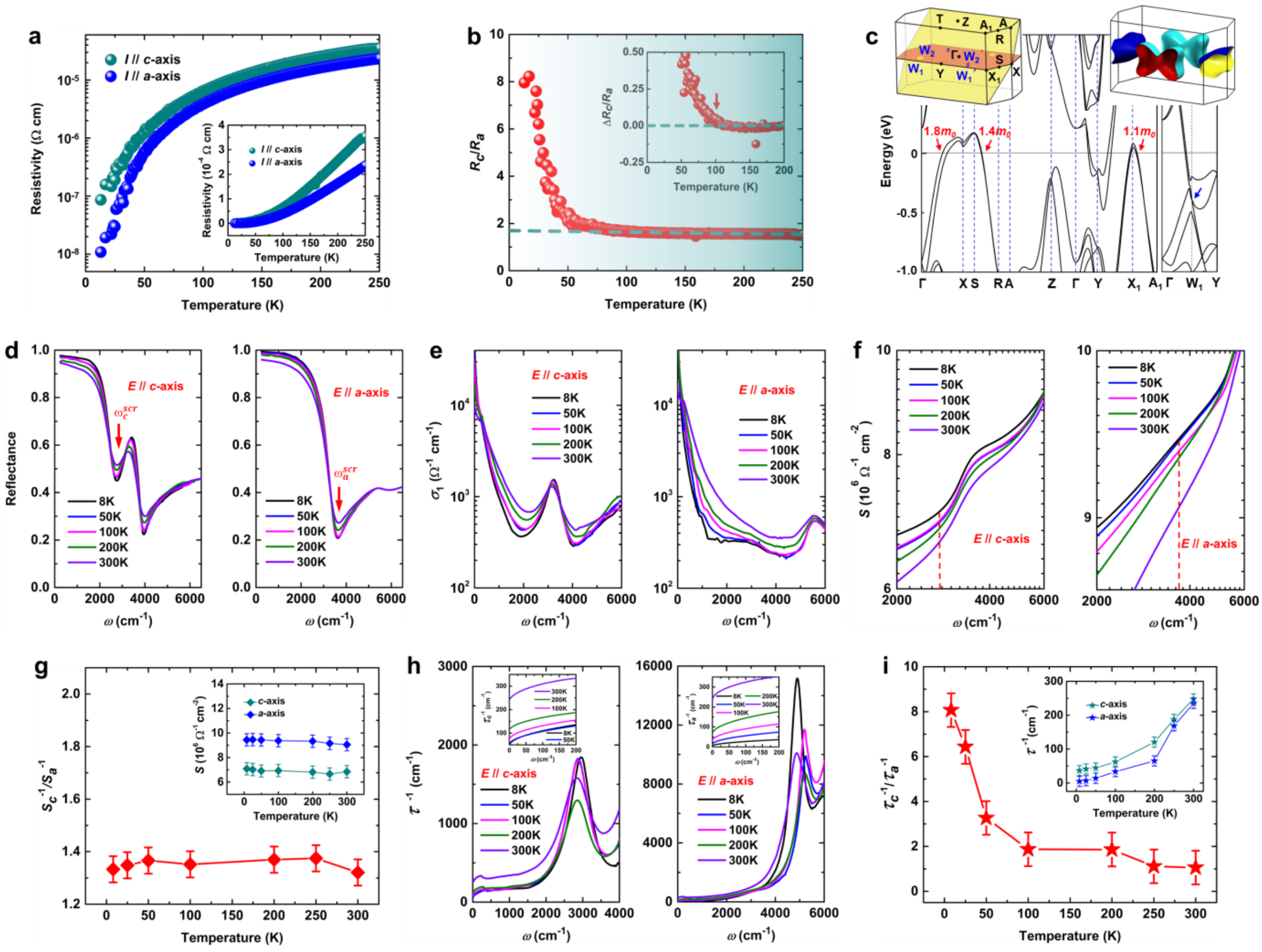

**Figure 4.** In-plane resistivity anisotropy and optical study of charge dynamics in the $WP_2$ single crystals. a) Temperature dependence of the resistivities on a logarithmic scale. The resistivities were measured with the electric currents (i.e., ***I***) applied along the *c*-axis and *a*-axis, respectively. The resistivities were plotted on the original scale in the bottom-right inset. b) Ratio between the *c*-axis and *a*-axis resistivities as a function of temperature (*T*). The inset displays the relative resistivity ratio $\Delta R_c/R_a$ obtained by subtracting the linear *T* dependence of the resistivity ratio (see the dashed line in b)). The red arrow indicates the onset temperature of an abrupt increase in the $\Delta R_c/R_a$. c) Calculated band dispersions along the high-symmetry lines of the Brillouin zone and near the Weyl point $W_1$. The upper left and right insets show four pairs of Weyl points in the Brillouin zone and the Fermi surfaces, respectively. d) Reflectance spectra measured with the electric field ***E*** // *c*-axis (the left panel) and ***E*** // *a*-axis (the right panel). The two red arrows in the left and right panels indicate the screened plasma frequencies $\omega_c^{src}$ and $\omega_a^{src}$ which correspond to the incident light with ***E*** // *c*-axis and ***E*** // *a*-axis, respectively. e) Real parts $\sigma_1$ of the optical conductivity with ***E*** // *c*-axis (the left panel) and *E* // *a*-axis (the right panel). f) Spectral weights (*S*) of the $\sigma_1$ with ***E*** // *c*-axis and ***E*** // *a*-axis. The two dashed vertical lines in the left and right panels are guides for eye showing the screened plasma frequencies. g) Ratio $S_c^{-1}/S_a^{-1}$ between the reciprocals of the Drude weights with ***E*** // *c*-axis and ***E*** // *a*-axis at different temperatures. The Drude weights $S_c$ and $S_a$ were plotted in the inset as a function of *T*. h) Scattering rate spectra $\tau_c^{-1}(\omega)$ and $\tau_a^{-1}(\omega)$ obtained with ***E*** // *c*-axis (the left panel) and ***E*** // *a*-axis (the right panel) at different temperatures. The insets display the magnified views of the $\tau_c^{-1}(\omega)$ and $\tau_a^{-1}(\omega)$. i) Ratio $\tau_c^{-1}/\tau_a^{-1}$ at different temperatures. The inset shows the *T* dependences of the $\tau_c^{-1}$ and $\tau_a^{-1}$.

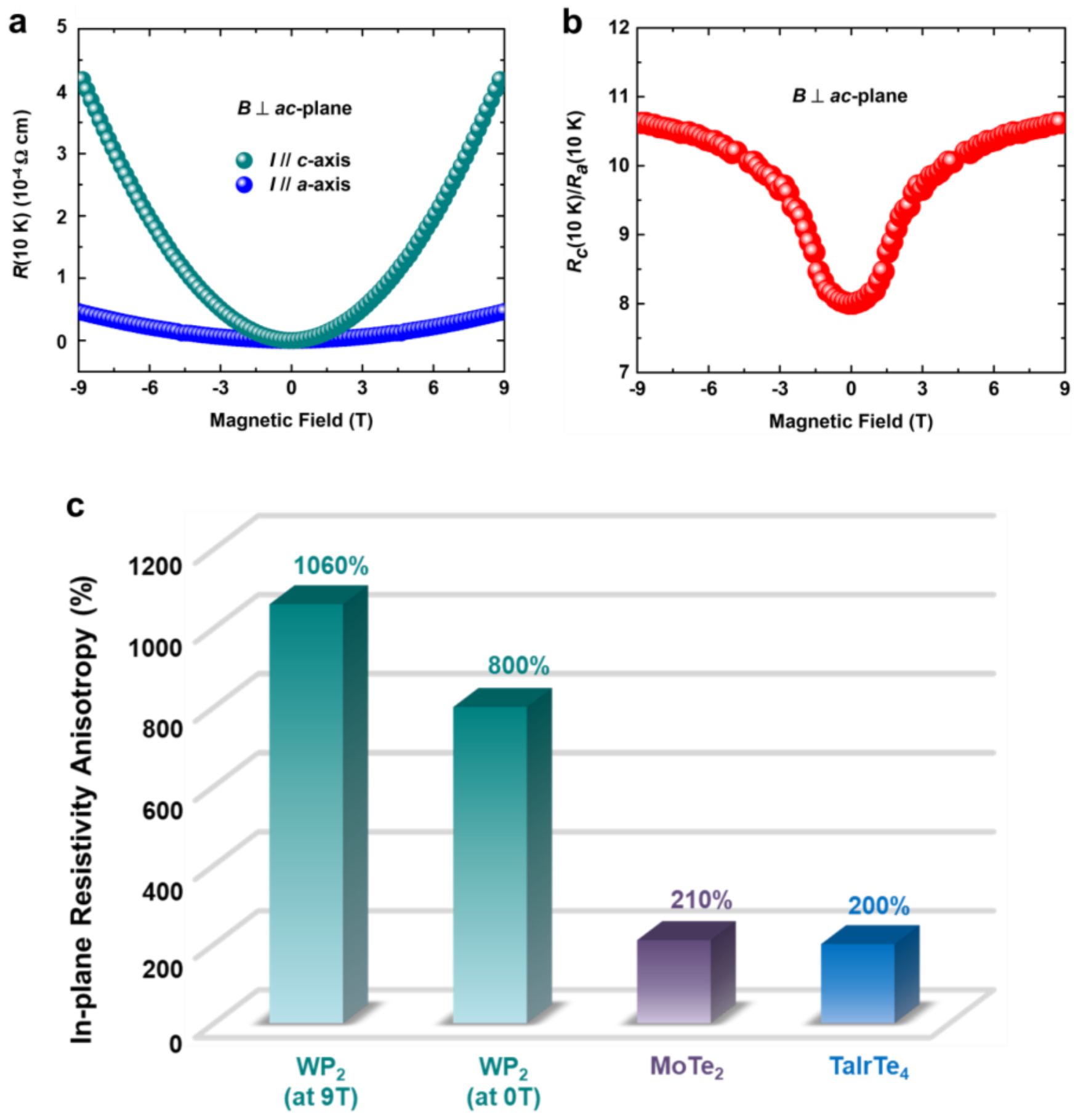

**Figure 5.** In-plane resistivity anisotropies of a type-II Weyl semimetal candidate $WP_2$ at magnetic fields. a) Magnetic-field-dependent resistivities measured at temperature $T$ = 10 K with the magnetic field (i.e., $B$) perpendicular to the *ac*-plane and the electric currents along the *a*-axis and *c*-axis. b) Ratio between the *c*-axis and *a*-axis resistivities at 10 K as a function of magnetic field. The ratio between the *c*-axis and *a*-axis resistivities at $T$ = 10 K goes up to ~ 10.6 at $B$ = 9 T. c) Reported in-plane resistivity anisotropies of the promising type-II Weyl semimetal candidates with their carrier densities comparable to $10^{21}$ $cm^{-3}$.

**Table 1.** The calculated Raman-active and infrared-active (IR) phonon modes, and experimental Raman-active phonon modes of $WP_2$ with their irreducible representations at room temperature.

| Symmetry | Activity | Calculated energy ($cm^{-1}$) | Experiment energy ($cm^{-1}$) |
|---|---|---|---|
| **$A_2$** | Raman | 157.4 | 159.2 ($P_1$) |
| **$B_2$** | Raman + IR | 166.8 | |
| **$A_1$** | Raman + IR | 168.4 | 168.4 ($P_2$) |
| **$B_1$** | Raman + IR | 250.3 | |
| **$A_2$** | Raman | 250.7 | 256.3 ($P_4$) |
| **$A_2$** | Raman | 275.8 | 282.2 ($P_5$) |
| **$A_1$** | Raman + IR | 277.4 | 284.1 ($P_6$) |
| **$B_2$** | Raman + IR | 284.3 | |
| **$B_1$** | Raman + IR | 311.7 | |
| **$A_1$** | Raman + IR | 341.8 | 359.9 ($P_8$) |
| **$B_2$** | Raman + IR | 346.8 | |
| **$A_1$** | Raman + IR | 381.0 | 392.4 ($P_9$) |
| **$B_2$** | Raman + IR | 432.1 | |
| **$A_1$** | Raman + IR | 498.2 | 511.7 ($P_{10}$) |
| **$B_2$** | Raman + IR | 525.7 | |